\documentclass{emulateapj}
\shorttitle{Dynamical Mass of GJ~802B}
\shortauthors{Ireland et al.}
\begin{document}

\title{Dynamical Mass of GJ~802B: a brown dwarf in a triple system}

\author{Ireland, M.J.}
\affil{Division of Geological and Planetary Sciences, California Institute of
  Technology, Pasadena, CA 91125}
\email{mireland@gps.caltech.edu}
\author{Kraus, A.}
\affil{Division of Physics, Mathematics and Astronomy, California Institute of
  Technology, Pasadena, CA 91125}
\author{Martinache, F., Lloyd, J.P.}
\affil{Cornell University, Ithaca, NY 14853, USA}
\author{Tuthill, P.G.}
\affil{School of Physics, University of Sydney, NSW 2006, Australia}

\begin{abstract}
 We report a dynamical measurement of the mass of the brown dwarf 
 GJ~802B using aperture-masking interferometry and astrometry. In
 addition, we report  
 the discovery that GJ~802A is itself a close spectroscopic
 non-eclipsing binary with a 19 hour period. We find the mass of
 GJ~802~B to be $0.063\pm0.005$M$_\sun$. GJ~802 has kinematics inconsistent
 with a young star and more consistent with the thick disk population,
 implying a system age of $\sim$10\,GYr. However, model
 evolutionary tracks for GJ~802B predict system ages of $\sim$2\,GYr,
 suggesting that brown dwarf evolutionary models may be
 underestimating luminosity for old brown dwarfs. 
\end{abstract}

\section{Introduction}

The boundary between stars and brown dwarfs
is defined as the mass at which the luminosity of old objects is
just dominated by hydrogen burning. This boundary is predicted to
occur at a mass between
0.07-0.072\,$M_\sun$ at [Fe/H]=0 and
$\sim$0.092\,$M_\sun$ at [Fe/H]=-3 \citep{Chabrier00,Burrows01}. 

However, this boundary, and the theoretical relationships that predict
effective temperature and luminosity as a function of mass and age,
are largely untested by observations. The observations that are
required to test these models are dynamical mass measurements of binary
and multiple-star systems, combined with accurate photometry and
distance determinations, preferably at known age. 
Many stars between 0.1 and 0.2\,$M_\sun$ now
have accurate dynamical masses \citep{Segransan00}, but objects with
masses between the hydrogen burning limit and 0.1\,$M_\sun$ so far do
not have accurate ($\la10$\%) mass and luminosity determinations.

Dependence on theoretical models has led to controversy about the
mass of brown dwarfs, particularly 
when objects are plausibly near the canonical planetary-mass boundary
of $\sim$13\,M$_{\rm jup}$ \citep[e.g.][]{Luhman07}. In the few cases where
the dynamical mass of a brown dwarf has been measured
\citep{ZapateroOsorio04,Stassun06}, precision is either inadequate to
truly constrain models, or the model fits have to make unusual
assumptions such as non-coeval systems \citep{Stassun07}.

GJ~802 is a M5 field dwarf system at $\sim$16\,pc that was
discovered to have a 
brown dwarf component through astrometry \citep{Pravdo05}. The
subsequent detection of the companion, GJ~802B, with aperture-masking
interferometry \citep{Lloyd06} made this an ideal target for dynamical
mass determination, due to the system's $\sim$3 year period. The high
contrast ($\sim$100:1) and small separation ($\la100$\,mas) of
GJ~802AB make this system too difficult for direct imaging
observations, but is in an ideal range for interferometric
detection. 

This paper reports aperture-masking interferometry detections of
GJ~802B at six epochs in three colors, enough to make good
measurements of the orbit, and the discovery that GJ~802A is itself a
close spectroscopic M dwarf 
binary. Section~\ref{sectObservations} describes the astrometric,
interferometric, spectroscopic and photometric observations that go
into the orbit determinations. Section~\ref{sectDiscussion} describes
the orbit of 
the close binary GJ~802Aab and Section~\ref{sectDynamMass} describes
the constraint on the dynamical mass of GJ~802B from 
the orbit of GJ~802AB. In Section~\ref{sectModelComp} we compare our results to
theoretical predictions and Section~\ref{sectConclusion} has a summary
and discussion.


\section{Observations}
\label{sectObservations}

The observations on which this paper is based consist of
seeing-limited astrometry, aperture-masking interferometry, infrared
and visible spectroscopy and photometry. The astrometry observations,
in addition to parallax and proper motion, provide the orbit of the
GJ~802Aab pair about the GJ~802AB center of mass. The aperture-masking
interferometry constrains the wide binary orbit by resolving
GJ~802B. The spectroscopy, in the infrared and visible, provides the
spectroscopic GJ~802Aab orbit, and the photometry is used to search for
eclipses of this inner pair. Each observation set will be discussed in turn.

\clearpage
\subsection{Astrometry Observations}

The astrometry observations in this paper come directly from
\citet{Pravdo05}. These observations were part of the Stellar Planet
Survey (STEPS) program, with observations made using a custom visible camera
mounted at the unfolded Cassegrain focus of the Palomar 200\,inch
telescope. As the raw astrometry was not published in that
paper and have not been made available to us, we extracted the
astrometric data from the figures in the paper, and then added the
motion due to the proper motion and parallax given in the paper's
table. These extracted values are presented in Table~\ref{tabAstro}, and 
include all epochs where there are at least 2 measurements given in
\citet{Pravdo05}. We give all of these epochs equal weight. The error
estimated from the scatter within individual epochs is 1.6\,mas,
however we assume errors of 1.7\,mas, to give a final fitted $\chi^2$
of 1.0 (see Section~\ref{sectDynamMass}). 


\begin{deluxetable}{llrr}
\tabletypesize{\scriptsize}
\tablewidth{0pt}
\tablecaption{Astrometry observation summary for GJ~802}
\tablecolumns{8}
\tablehead{\colhead{Date (UT)} & \colhead{JD-2450000}
& \colhead{$\Delta$RA (mas)} & \colhead{$\Delta$Dec (mas)}}
\startdata
1998-07-06 &  1000.75 &      0.00 &      0.00 \\
1999-07-20 &  1379.75 &    901.50 &   1806.99 \\
1999-09-23 &  1444.75 &   1002.89 &   2092.99 \\
2000-07-13 &  1738.75 &   1780.05 &   3510.93 \\
2001-07-02 &  2092.75 &   2631.23 &   5150.19 \\
2002-07-07 &  2462.75 &   3513.06 &   6908.15 \\
2003-09-09 &  2891.75 &   4496.65 &   8932.57 \\
2004-07-12 &  3198.75 &   5276.92 &  10367.24 \\
2004-09-15 &  3263.75 &   5371.75 &  10654.40 \\
\enddata
\label{tabAstro}
\end{deluxetable}

\subsection{Aperture-masking interferometry}
\label{sectApertureMasking}

The technique of non-redundant aperture masking has been well-established
as a means of achieving the full diffraction limit of a single telescope
\citep[e.g.][]{Michelson20,Baldwin86,Tuthill00}. It involves placing
a mask with an array of holes of non-redundant spacing in the
pupil-plane of a telescope, and analyzing the recorded images as
interference fringes on a number of discrete baselines. Recently, our group
has used this technique behind adaptive optics at the Palomar 200 inch
and Keck telescopes to increase the magnitude limit of
the technique beyond that achievable in a seeing-limited speckle
mode. The reason for the technique's success over direct
adaptive-optics imaging is that the calibration is independent of
structure of the wavefront over scales larger than a 
single sub-aperture, but it still preserves the angular resolution of
the full aperture.

Aperture-masking observations of GJ~802 were
made using both the PHARO camera of the Palomar 200 inch telescope
\citep{Hayward01} and
the NIRC2 camera of the Keck II telescope. The Palomar aperture-masking
mode of the PHARO camera is described in \citet{Lloyd06}. The NIRC2
aperture-masking mode is similar, although the masks are placed in the
filter wheels of the NIRC2 camera rather than a designated pupil wheel
as is the case with PHARO. The masks create fringe patterns
in the image-plane of each detector. Our primary observable,
closure-phase, is extracted from these fringe patterns and models are
fit to the closure phases. A summary of all aperture-masking
observations, along with 
the detected binary properties, are given in Table~\ref{tabApMasking}.

For the NIRC2 and PHARO aperture-masking experiments, both 9 and 18
hole masks are available. The 9 hole mask results in 36 simultaneous
baselines and 28 
independent closure-phases. The 18 hole mask results in 153
simultaneous baselines and 136 independent closure-phases, but has
half the throughput of the 9 hole mask and spreads the light over 4
times as many pixels as the 9 hole mask.

All observations were done with a 9 hole mask in H and K bands, except the
June 2007 J$_{\rm cont}$ observation for which a 18 hole mask was used. 
In the absence of sky rotation, the field-of-view of the aperture-masking
experiment is given by $\lambda/2B_S$, where $B_S$ is the shortest
baseline in the mask. Outside this field-of-view, the position angle
and separation become ambiguous without multiple exposures at
different sky rotations. Although the 9 hole mask was well suited to H
and K observations at both Keck and Palomar, the $\sim$73\,mas separation of
GJ 802B would have been outside the nominal 9 hole mask field of view
in J band. 

In a similar manner, the aperture masking inner working angle is given
by $\lambda$/2B$_L$, where B$_L$ is the longest baseline (near to the
full aperture 
size). At this separation, the maximum closure phase signal is equal to
the contrast ratio of the binary (i.e. 0.01 radians or 0.6 degrees for a 100:1
binary). In the case of the PHARO observations of June 2004 and
October 2006, these nominal inner detection limits are 43 and 57\,mas
respectively. Near this inner edge, the contrast ratio and separation
for model-fitting are degenerate. Despite their large errors on
separation, these detections are therefore reliable. We
explicitly list the correlations between separation and contrast in
Table~2.

We have made several improvements to the analysis pipeline and
observing procedure since the
original detection published in \citet{Lloyd06}. The most important
improvement has been in the observations themselves: we have
generally been much more careful in using calibrators of similar
brightness, colors and position in the sky. This procedure was not
used carefully for the discovery epoch of September 2004, and we find
that the astrometry depends on which of several unsuitable calibrators
are used in the analysis. Therefore, we choose not to use this epoch
in our analysis here, and it is not presented in Table~\ref{tabApMasking}. 

We have windowed the data prior to Fourier-transforming with a tighter
function than previous publications to minimize residual chip-based effects 
(e.g. of bad pixels). The window size is of the form exp$(-r^4)$
with a full-width half-maximum of $1.6\lambda/D_H$, where $D_H$ is our
aperture-mask hole diameter projected onto the primary mirror. 
This tighter window also has the effect of
spatially-filtering the interferometric data. The raw frames from the
PHARO camera are taken in a mode where all (non-destructive) reads are
saved. This allows us to split the data in post-processing into a
variety of integration times, 
where smaller integration times have less atmospheric and AO system
noise but more readout noise. We now carefully choose an optimal
number of sub-reads to analyze data of a given brightness: for GJ~802
we split the data into sub-frames spaced by 2 reads, which have
integration times of 862\,ms each.


\begin{deluxetable*}{llllrrrr}
\tabletypesize{\scriptsize}
\tablewidth{0pt}
\tablecaption{Aperture-masking observation summary for GJ~802}
\tablecolumns{8}
\tablehead{\colhead{Date (UT)} & \colhead{JD-2450000}
& \colhead{Instrument} & \colhead{Filter}
& \colhead{Separation}& \colhead{Position Angle} &
  \colhead{Contrast\tablenotemark{a}} & \colhead{Sep./Contrast} \\
 & & & 
& \colhead{(mas)}& \colhead{(degrees)} &
  \colhead{(B/A)} & \colhead{correlation}}
\startdata
2004-06-06 & 53163.0 &  PHARO &      H &  55.1$\pm$ 21.4 &  16.4$\pm$  6.8 & 0.013$\pm$0.006 &  0.87 \\
2006-06-23 & 53910.0 &  NIRC2 &     Ks &  85.8$\pm$  3.5 & 200.9$\pm$  2.2 & 0.015$\pm$0.003 &  0.09 \\
2006-10-10 & 54018.7 &  PHARO &      H &  94.6$\pm$  9.7 & 207.6$\pm$  7.4 & 0.009$\pm$0.003 & -0.13 \\
2006-10-10 & 54018.7 &  PHARO &     Ks &  81.1$\pm$ 24.5 & 196.7$\pm$  4.9 & 0.011$\pm$0.003 &  0.75 \\
2007-05-31 & 54252.0 &  PHARO &   CH4S &  77.4$\pm$ 10.3 &  24.1$\pm$  5.2 & 0.009$\pm$0.002 & -0.26 \\
2007-06-05 & 54257.0 &  NIRC2 &  Jcont &  73.3$\pm$  2.9 &  21.5$\pm$  2.1 & 0.008$\pm$0.002 &  0.04 \\
2007-06-06 & 54258.0 &  NIRC2 &     Kp &  72.6$\pm$  1.7 &  19.2$\pm$  1.2 & 0.010$\pm$0.001 &  0.01 \\
2007-07-31 & 54312.9 &  PHARO &      H &  86.2$\pm$  5.8 &  24.0$\pm$  3.5 & 0.010$\pm$0.002 & -0.40 \\
2007-08-29 & 54341.8 &  PHARO &      H &  85.4$\pm$  6.0 &  20.2$\pm$  3.2 & 0.011$\pm$0.002 & -0.44 \\
\enddata
\label{tabApMasking}
\end{deluxetable*}

\clearpage
\subsection{Infrared Spectroscopy}
\label{sectSpectroAstro}

The NIRC2 camera at the Keck II telescope was used on UT 2006-08-10 
in a spectroscopic mode for the purpose of attempting to detect the
spectrum of GJ~802B directly. The observations were made in the H-band
with a bandpass from 1.52 to 1.62\,microns. We used a grism with a
resolving power of 17580 per pixel and a 4 pixel wide slit, giving a
spectral resolution of $\sim$4000. These observations did not succeed in
their primary purpose, mainly because poor seeing forced us to use H
instead of J-band, where the spectral differences between a mid-L and
mid-M dwarf were not large enough to give a clearly detectable GJ~802~B 
signal in the data.

It was noticed, however, that lines in the spectra of GJ~802 were
double. To calibrate the spectra, we first used Xe and Ar lamps to
calibrate the wavelength scale, which was fitted with a second-order
polynomial. We calculated the model atmospheric and grism transmission by taking
spectra of the F8V star HD~136118 and dividing by a template spectrum
of HR~4375 (a G0V star) from \citet{Meyer98}.
In reducing these spectra to their two components, GJ~777B was
used as the spectral-type standard and the radial velocity standard, 
as its wide companion GJ~777A has a precise radial velocity of
-45.350$\pm$0.004\,km\,s$^{-1}$ \citep{Naef03}, and should be within
1\,km\,s$^{-1}$ from GJ~777B's radial velocity due to potential
orbital motion. As a 
check of our calibration procedure, we used Xe and Ar lamp
calibration of the NIRC2 grism wavelength scale to measure the radial
velocity of GJ~777B based on two Al I and one Ca I lines to be
$-41\pm5$\,km\,s$^{-1}$, consistent with GJ~777A's radial velocity.

Least square fits were made to the continuum subtracted spectra of GJ 802, 
based on a model made up of the sum of two shifted GJ 777B
spectra. We chose this technique rather than a cross-correlation
because the two spectra were not separated well enough to give clearly
separate peaks in the cross-correlation. The errors in the velocity
difference come straight from the least-squares
fitting process, where the uncertainty in the GJ~802 spectrum was set
to 1.3\% of the mean flux so that the model fit had a reduced $\chi^2$
of 1. An uncertainty on 
the velocity difference based on deviations from a linear fit to the
measured velocities would be 2.5\,km\,s$^{-1}$. The results
of this fitting process are given in Table~\ref{tabInfraredRV}. The
magnitude difference in $H$-band derived from the fit is zero within
errors, which suggests that the components have nearly equal masses.

It is difficult to estimate the uncertainties on
the absolute velocity calibration (not the primary purpose of
our observations), so we assign
an error of 5\,km\,s$^{-1}$ based on 
the absolute calibration of GJ~777B's radial velocity against the Xe
and Ar lamps. There is good
reason to expect that the absolute calibration of these observations
is significantly worse than the relative calibration between the Aa
and Ab spectra, because the point spread function delivered by the AO
system was not guaranteed to be well-centered on the slit. However, as
both stars had the same point spread function, with their maximum
possible spatial separation  of 0.7\,mas corresponding to
$\sim$1\,km\,s$^{-1}$, the relative calibration between Aa and Ab
spectra should be good compared with the 2\,km\,s$^{-1}$ statistical
error in the velocity difference.  

\begin{deluxetable}{lrrrr}
\tabletypesize{\scriptsize}
\tablewidth{0pt}
\tablecaption{Infrared Radial Velocities for GJ~802Aab}
\tablecolumns{5}
\tablehead{\colhead{JD- 2450000} & \colhead{RV$_{Ab}$} & \colhead{RV$_{Aa}$} &
  \colhead{$\Delta$RV} & \colhead{H$_{Aa}$-H$_{Ab}$\tablenotemark{a}}}
\startdata
3956.9219 &   -6.9$\pm$5.0 &  -93.5$\pm$5.0 &  -86.6$\pm$2.0 &  0.15$\pm$0.06 \\
3956.9336 &  -13.2$\pm$5.0 &  -86.1$\pm$5.0 &  -72.9$\pm$2.0 &  0.19$\pm$0.06 \\
3956.9375 &  -12.9$\pm$5.0 &  -83.7$\pm$5.0 &  -70.9$\pm$2.0 &  0.24$\pm$0.06 \\
3956.9453 &  -16.3$\pm$5.0 &  -83.6$\pm$5.0 &  -67.3$\pm$2.0 & -0.03$\pm$0.07 \\
\enddata
\label{tabInfraredRV}
\tablenotetext{a}{Difference in apparent H-band magnitudes as derived from
the fit of two shifted GJ~777B spectra. See text.}
\end{deluxetable}

\subsection{High-Resolution Optical Spectroscopy}
\label{sectVisibleSpectroscopy}

We obtained multi-epoch spectra of GJ 802 using the East-Arm Echelle
(EAE) on the Palomar Hale 5m telescope. The EAE is a high-resolution
spectrograph, capable of achieving spectral resolution of
$R\sim30000$. Most of our observations span a wavelength range of
4000-10000 $\AA$, but the observations taken in December 2006
(before official commissioning) span a significantly bluer range
(3500-8000 $\AA$). Most of our spectra were binned in the spectral
direction in order to reduce read noise, so the effective
resolution for our observations was $R\sim$20000. Since GJ 802 is very
red, none of our spectra have any useable signal short-ward of
$\sim$5500 $\AA$.

We observed GJ 802 at four epochs on 2006 December 15 and six more
epochs spanning 2007 May 06-08. We also observed the late-type stars
GJ 51 (in December) and GJ 581, GJ 686, and GJ 699 (in May) as
spectral type standards. All observations of GJ~802 used integration
times of 300-900s, while our standard star observations used a wider
range of integration times since some were significantly brighter. 

\begin{figure}
\epsscale{.80}
\plotone{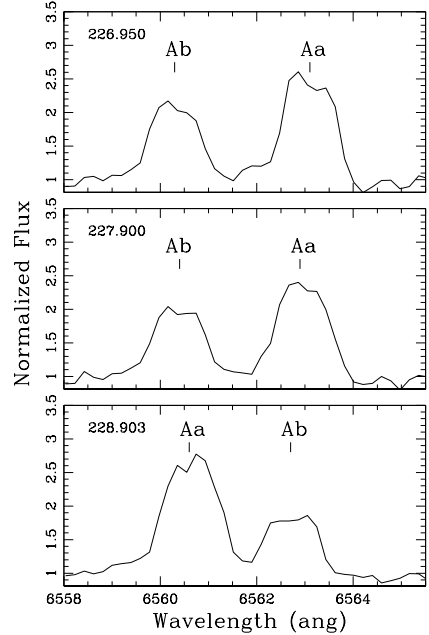}
\caption{High-resolution spectra for GJ 802 centered on the H$\alpha$
  emission line. Both components show resolved H$\alpha$ emission, but
  the H$\alpha$ emission line strengths appear to be variable on a
  time-scale of days.}
\label{figVisSpectra}
\end{figure}

We bias-subtracted, flat-fielded, and extracted our spectra using
standard IRAF tasks. Wavelength calibration was achieved with respect
to a Thorium-Argon lamp that was observed at the beginning of each
night; preliminary tests suggest that telescope flexure results in
wavelength calibration variations of no more than $\sim$0.015 $\AA$
($\sim$0.5 km s$^{-1}$) in the vicinity of the H$\alpha$ emission line
at 6563 A. In Figure~\ref{figVisSpectra}, we show a representative
segments of several spectra for GJ~802 around the $H\alpha$ wavelength
range. These spectra demonstrate the double-lined nature of GJ~802; this
plot also shows that the H$\alpha$ emission line strengths are not
constant, but vary on a time-scale of days.

All of our science spectra exhibit H$\alpha$ emission, so we have
directly determined the component radial velocities from the line
centroids. For epochs where the lines were not clearly resolved, we
fit the spectra with a pair of Gaussian emission lines with the same
FWHM as the resolved measurements. We list all of these radial
velocities, including heliocentric corrections, in 
Table~\ref{tabVisRV2}; we also list our observed radial velocity
standards in Table~\ref{tabStandardRV}. The mean radial velocity of
the standards was $3.0$\,km\,s$^{-1}$ higher than the radial
velocities derived from the Thorium-Argon lamp calibration, with a
dispersion of 1.7\,km\,s$^{-1}$ (standard error on the mean
0.5\,km\,s$^{-1}$). We have subtracted this offset from all radial
velocities reported in Table~\ref{tabStandardRV} and
Table~\ref{tabVisRV2}.

We did not use any cross-correlation techniques to determine overall
RV fits because we are still attempting to characterize the wavelength
and flux calibrations of the instrument, including limits on potential
variability; given the long time baseline of our dataset, any
improvement in the precision would yield only very minor improvements
in the orbital ephemerides. 

\begin{deluxetable}{lrrr}
\tabletypesize{\scriptsize}
\tablewidth{0pt}
\tablecaption{High-Resolution Spectroscopy of GJ~802. }
\tablehead{\colhead{JD - 2450000} & 
\colhead{ RV$_{Ab}$} & \colhead{ RV$_{Aa}$} & \colhead{$\Delta$ RV}
}
\startdata
  4084.57847 & -107.7$\pm$ 5.0 &   10.2$\pm$ 5.0 & -117.9$\pm$ 2.8 \\
  4084.59097 & -111.8$\pm$ 5.0 &   15.7$\pm$ 5.0 & -127.5$\pm$ 2.8 \\
  4084.60347 & -118.2$\pm$ 5.0 &   18.0$\pm$ 5.0 & -136.2$\pm$ 2.8 \\
  4084.70625 & -116.9$\pm$ 6.0 &   19.4$\pm$ 6.0 & -136.3$\pm$ 4.2 \\
  4084.71389 & -118.7$\pm$ 6.0 &   18.5$\pm$ 6.0 & -137.2$\pm$ 4.2 \\
  4084.72014 & -113.2$\pm$ 7.0 &   16.6$\pm$ 7.0 & -129.8$\pm$ 7.1 \\
  4226.94971 & -102.3$\pm$ 2.0 &   23.2$\pm$ 2.0 & -125.5$\pm$ 2.8 \\
  4227.90039 &  -99.3$\pm$ 2.0 &   17.9$\pm$ 2.0 & -117.2$\pm$ 2.8 \\
  4228.84473 &  -21.4$\pm$ 2.0 &  -59.8$\pm$ 2.0 &   38.4$\pm$ 2.8 \\
  4228.90332 &    9.7$\pm$ 2.0 &  -86.7$\pm$ 2.0 &   96.4$\pm$ 2.8 \\
\enddata
\label{tabVisRV2}
\end{deluxetable}

\begin{deluxetable}{lrrr}
\tabletypesize{\scriptsize}
\tablewidth{0pt}
\tablecaption{High-Resolution Spectroscopy of Standards. }
\tablehead{\colhead{Star} & \colhead{JD - 2450000} & 
 \colhead{RV} & \colhead{Literature RV} }
\startdata
GJ 51 &
  4084.66134 & -10.9$\pm$3  & -7.7\tablenotemark{a}  \\
GJ 393 &
  4226.70127 & +4.6$\pm$3 & +8.3\tablenotemark{b} \\
GJ 476  &
 4226.75556 & +29.6$\pm$3 & +31.6\tablenotemark{a} \\
GJ 686 &
 4226.83443 & -11.4$\pm$3 & -9.5\tablenotemark{b}  \\
GJ 476 &
 4227.73785 & +30.6$\pm$3 & +31.6\tablenotemark{a} \\
GJ 514 &
 4227.74175 & +11.2$\pm$3  & 14.6\tablenotemark{b}\\
GJ 526 &
 4227.74530 & +13.0$\pm$3  & 16.0\tablenotemark{a} \\
GJ 581 &
  4227.74907 & -13.8$\pm$3  & -9.4\tablenotemark{b}  \\
GJ 393 &
 4228.63097 & +4.9$\pm$3  & +8.3\tablenotemark{b} \\
GJ 387.2 &
 4228.64279 & -16.8$\pm$3 & -16.5\tablenotemark{a} \\
GJ 699 &
 4228.82439 & -116.9$\pm$3   & -110.5\tablenotemark{b} \\
\enddata
\tablenotetext{a}{\citet{Gizis02}}
\tablenotetext{b}{\citet{Nidever02}}
\label{tabStandardRV}
\end{deluxetable}

\subsection{Photometry}
\label{sectPhotometry}

Given that the GJ~802Aab binary had such a short period, and
preliminary analysis of the orbit suggested that
inclination was possibly high, we used the Palomar 60-inch robotic
telescope to search for eclipses. The Palomar 60-inch telescope has a
single instrument, a CCD imager with an 11$\times$11 arc-minute
field-of view and a selection of broad and narrow-band filters. We
took series of images in  $g$, $i$ and H-$\alpha$
filters around the predicted times of eclipses based on the
radial-velocity orbit (when the velocity difference was zero). 
The H-$\alpha$ filter was used to search for an eclipse of
the chromospheric emission. 

The observations in $g$ and $i$ filters spanned 30 minutes on either side of a
predicted eclipse at Julian day 2454311.889, and the observations in
H-$\alpha$ spanned 30 minutes on either side of a predicted eclipse at
Julian day 2454315.870. These eclipse times were based on a
preliminary orbital fit (see Section~\ref{sectDiscussion}). As the
radial velocity amplitudes of Aa and Ab 
were roughly equal (see the velocities in Table~\ref{tabVisRV2}), we
can assume that the masses are roughly equal and that the primary and
secondary eclipses would have roughly equal depths.

We computed aperture photometry for GJ~802 on the calibrated images,
using the median photometric variation of 10 nearby field stars to
compute the sky transmission in each frame. No eclipses 
were found in any filters. For the $g$ and $i$ filters, the RMS
photometry scatter was 0.01 and 0.003\,mag respectively with sampling at
1.5\,minute intervals. For the H-$\alpha$ filter, the scatter was
0.01\,mag, and the sampling at 1\,minute intervals. The excess in our
filter due to the H-$\alpha$ emission was 22.7\% of the continuum,
calculated by comparison with photometry from an off-line narrow-band
filter and consistent with our spectroscopic observations. From these
data we can place 2$\sigma$ upper limits for a continuum eclipse depth
of 0.005\,mag and for a H-$\alpha$ eclipse depth of 0.01\,mag.



\section{Orbit of the close pair GJ~802A}
\label{sectDiscussion}

\begin{figure}
 \plotone{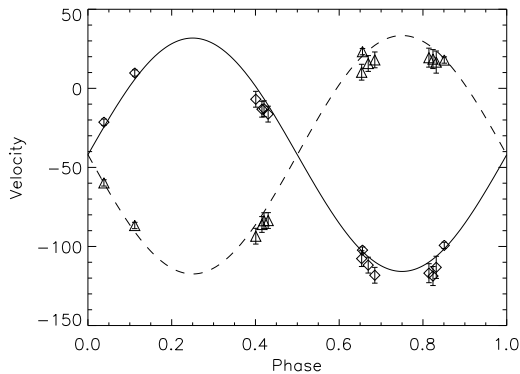}
 \caption{Best fit orbit for GJ~802Aab, showing the $Aa$ velocities
   (solid line, diamonds) and the $Ab$ velocities (dashed line, triangles).}
\label{figRVAB}
\end{figure}

\begin{figure}
 \plotone{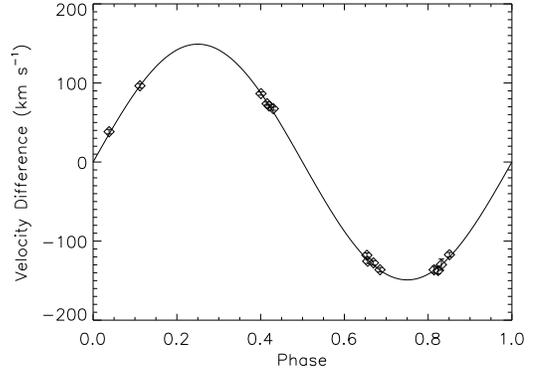}
 \caption{Best fit orbit for GJ~802Aab, showing the difference in
 velocities between the two components.}
\label{figRV}
\end{figure}

We chose to fit the radial velocities of GJ~802Aab with a circular orbit only,
because the 19 hour period of the system is 10 times shorter
than the canonical cutoff of 7 days for tidal circularization in
low-mass main-sequence stars \citep{Zahn89}. Figure~\ref{figRVAB}
shows the best fit equal-mass circular orbit with the measured radial
velocities, and  Figure~\ref{figRV} shows the best fit circular orbit
with the measured radial velocity differences. 
We made this fit by first examining by eye the 2007 May radial
velocities, concluding that the 
orbital semi amplitude was between 135 and 155 km\,s$^{-1}$, with a
period of approximately 0.8\,days. We then fit to the absolute value
of the velocity difference by using a grid search. We searched the
range of 0.7 to 0.9 days with $10^{-5}$ days spacing for period, the
full range of epoch (modulo half a period) with $10^{-3}$ days
spacing, and total velocity amplitudes of the two components $K1 +
K2$, between 135 and 155\,km\,s$^{-1}$ with
4\,km\,s$^{-1}$ spacing. The reduced $\chi^2$ for the final fit is
0.70 with 11 degrees of freedom. No other (aliased) minima have
reduced $\chi^2$ values less than 6. We assigned signs to the radial
velocity differences only after finding this fit, as GJ~802Aab is
so close to equal brightness that it was difficult to tell which spectrum
was GJ~802Aa and which was GJ~802Ab.

Best fit orbital parameters are given in Table~\ref{tabOrbitAab}. It
is not clear from these observations which component is the more
massive, as the mass ratio $M_{Ab}/M_{Aa} (= K1/K2)$ is $0.98\pm0.03$.
This ambiguity is seen in Figure~\ref{figRVAB}. The
mass ratio and center of mass radial velocity are constrained best by
the May 2007 observations, which are best calibrated and where the 
two spectra swap their positions, and the absolute calibration of the
radial velocities is most certain. These data have significantly
smaller errors in Figure~\ref{figRVAB} and are at phase $\sim$0.03,
0.1, 0.65 and 0.85. Including a $\sim$3\,km\,s$^{-1}$
uncertainty for the orbital motion of GJ~802Aab with respect to the
GJ~802AB center of  mass, we obtain a radial velocity for the system
of -42$\pm$4\,km\,s$^{-1}$. 

\begin{deluxetable}{lllrrrrrr}
\tabletypesize{\scriptsize}
\tablewidth{0pt}
\tablecaption{Radial Velocity Solution for GJ~802Aab}
\tablecolumns{2}
\tablehead{\colhead{Parameter} & \colhead{Value}}
\startdata
Epoch (HJD) & 2454140.530$\pm$0.001 \\
Period (d)  & 0.795340$\pm$0.000003 \\
K1 + K2  & 149.1$\pm$1.5 km\,s$^{-1}$\\
K1       & 73.8$\pm$1.4 km\,s$^{-1}$\\
K2       & 75.4$\pm$1.4 km\,s$^{-1}$\\
(M$_{Aa}$+M$_{Ab}$)sin$^3$i  & 0.273$\pm$0.008 M$_\sun$\\
a sin(i)  & 2.343$\pm$0.024 R$_\sun$ \\
Inclination      & 77$<$i$<$83 \\
\enddata
\label{tabOrbitAab}
\end{deluxetable}

The inclination limit in Table~\ref{tabOrbitAab} comes from the
following analysis of both the lack of eclipses and a model-dependent
mass estimate for the system.  
Based on a stellar radius of 0.16\,$R_\sun$,
predicted by the models of \citet{Baraffe98} for components of mass
$\sim$0.14\,M$_\sun$ (applicable to an edge-on system with 
equal-mass components and (M$_{Aa}$+M$_{Ab}$)sin$^3$i from Table~\ref{tabOrbitAab}), 
the eclipse would
have lasted a maximum of 50 minutes. For a grazing 
eclipse with 5.8\% of one star eclipsed, the eclipse would have lasted
25 minutes: well within our 60 minute observing window (Section~\ref{sectPhotometry}). 
A grazing eclipse 
occurs at a projected separation of
twice the stellar radius, or $\sim$0.32\,$R_\sun$.  
As no eclipse occulting even 1\% of the surface of one of the stars 
was found in any filter (Section~\ref{sectPhotometry}),
the inclination of GJ~802Aab is limited to $<83$ degrees. Given that 
(M$_{Aa}$+M$_{Ab}$)sin$^3$i = 0.273$\pm$0.008, this in turn limits the
total mass of  GJ~802A to $>0.279\pm0.008$.

The models of \citet{Baraffe98}
have been verified to predict the K-band mass-luminosity relationship
for field dwarfs correctly to within $\sim$5\% in
mass for dwarfs of mass greater than
0.1\,$M_\sun$\citep{Delfosse04}. For stars of mass
$\sim$0.11-0.18\,$M_\sun$, this relationship is independent of age at
the 1\% level for ages between 0.5 and 10\,Gyr. This age range is
applicable to the likely GJ~802 age of $\ga$6\,Gyr, based on its
activity \citep{Pravdo05} and kinematics
(Section~\ref{sectModelComp}). This 
relationship is also relatively steep: at a mass of 0.13\,M$\sun$, a 6\%
error in K-band flux from the distance of \citet{Pravdo05} translates
to only a 2.5\% error in mass. Therefore, the uncertainty in the mass
is likely dominated by the K-band mass-luminosity relationship (at
least until the models are tested at higher precision), and we assign
a standard deviation of 5\% to the model-predicted mass.
These models predict masses for GJ~802Aa and GJ~802Ab of
0.134\,$M_\sun$, using a parallax of 63\,mas \citet{Pravdo05} and
2MASS photometry. In
turn, this places a maximum-mass constraint on 
GJ~802Aab at $\sim2\sigma$ of 0.295\,M$_\sun$, limiting the
inclination of GJ~802Aab to between 77 and 83 degrees. Note that the
$K$-band luminosity - mass relationship is little affected by
metallicity, changing by $<1$\% if we were to use the models of
\citet{Montalban00} at [Fe/H]=-1.



\section{Dynamical mass of GJ~802B}
\label{sectDynamMass}
The full solution for the GJ~802AB orbit requires 16 parameters: 5 for
parallax and proper motion, 7 for the orbital solution of B with
respect to Aab, 1 for the photo-center semi-major axis, and 3 for the
contrast of B with respect to Aab in J, H and K bands. The reason that
contrast had to be added in to the orbital solution was that many of
the parameters in Table~\ref{tabApMasking} had strong degeneracies. To
fully explore this large parameters space around the best-fit
solution, we used a Markov Chain Monte Carlo technique
\citep[e.g.][]{Bremaud99}, a method that has often been used in 
astronomy for cosmological parameter estimation
\citep[e.g.][]{Knox01}. Key advantages of this 
technique are the ability to easily include the covariance matrix of
the data (in our case the aperture-masking fits), and the ability to
easily calculate the posterior probability function of derived
parameters, such as the mass of GJ~802B. 

The probability that a particular set of parameters is contained in
the final chain is proportional to the likelihood function, which is
proportional to $\exp(\chi^2/2)$. Due to the correlations between
derived parameters from aperture-masking, $\chi^2$ is not just the sum of
normalized deviates, but makes use of the covariance matrix of the data:

\begin{equation}
 \chi^2 = ({\bf m(\theta)}-{\bf d})^tC^{-1}({\bf m(\theta)}-{\bf d}).
\end{equation}

Here ${\bf d}$ is the vector of data values, ${\bf m}$ is the model
for these data based on parameters ${\bf \theta}$, $C$ is the data
covariance matrix and $^t$ represents a transpose. Covariances between
different epochs were assumed 
to be zero: only the derived separation, position angle and contrast
for a single epoch of aperture-masking data had non-zero covariances.

We used a Markov Chain of length $1.4\times 10^5$, with a $1.4\times 10^4$
burn-in time. The best orbital solution had a reduced $\chi^2$ of
1.0 because we chose the error in the STEPS data to be 1.7\,mas in
each axis. As in \citet{Pravdo05}, we used a $2\pm1$\,mas conversion
from relative to absolute parallax.

We first conducted an exploratory unconstrained fit to the data, but
found that the errors on parallax
and orbital semi-major axis were too large to give a useful total
mass. 
This fit with single parameter
1$\sigma$ confidence limits is given in the first column of
Table~\ref{tabOrbit}, 
and has a mass for GJ~802Aab that is too high by $\sim$1.1$\sigma$
from the mass derived from the absolute magnitude of GJ~802Aab. 
Therefore, we fixed fixed the masses of GJ~802Aa and GJ~802Ab to be  
0.134\,$M_\sun$, based on their absolute $K$-magnitudes and the models
of \citet{Baraffe98} (see Section~\ref{sectDiscussion}). Note that the
same value of parallax of 63\,mas was used in both
Section~\ref{sectDiscussion} and here in calculating the K-band
absolute magnitude. This fit is
given in the central column of Table~\ref{tabOrbit}. 

\begin{figure}
 \plotone{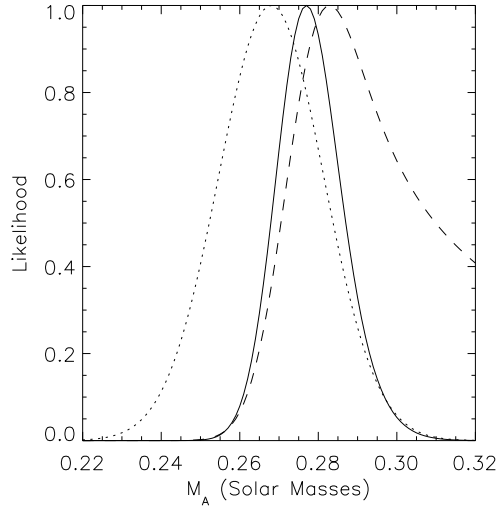}
 \caption{The apriori constraint on the mass of GJ~802A, from the
   K-band luminosity (dotted line), the radial velocity orbit (dashed
   line), and the joint likelihood (solid line).}
\label{figMassLikelihood}
\end{figure}

Finally, we note that we can not really ``fix'' the mass of GJ~802A as
it has an uncertainty. We therefore calculated an apriori constraint
on the mass of GJ~802Aab  
using the following: the mass-K magnitude relationship
with its assumed 5\% RMS error (see Section~\ref{sectDiscussion}); 
the value of $M$sin$^3(i)$ from
Table~\ref{tabOrbitAab}; and an orbital orientation assumed random with
$i<83$ degrees. Together, these constraints give a mass of
$0.2775\pm0.0082$ for the mass of GJ~802Aab. The assumed likelihood
curve resulting from these two constraints is given in
Figure~\ref{figMassLikelihood}: it can be seen that a Gaussian
likelihood curve is a good approximation for our prior knowledge of
$M_A$. The knowledge of this mass was added to 
the Markov-Chain Monte Carlo as an apriori constraint to give our
preferred values in the final column of Table~\ref{tabPhoto}. The
final GJ~802B mass of 
0.063$\pm$0.005\,$M_\sun$ is the most accurate mid-late L-dwarf mass
reported in the literature so far.

Although the Markov-Chain Monte Carlo technique is an excellent technique for
accurately calculating posterior probability distributions of derived
parameters, it can be difficult to intuitively understand the
magnitude of derived uncertainties from the chain output. Therefore,
we will examine the origin of the 8\% uncertainty on the GJ~802B mass
individually. The mass of the $B$ component directly relates to the
mass of the $A$ components by the ratio of astrometric and orbital
semi-major axes $a$ and $\alpha$ \citep[e.g][]{McCarthy88}:

\begin{equation}
 M_B = \frac{\alpha/a}{1 - \alpha/a}M_A.
 \label{eqnMBIntuitive}
\end{equation}

Therefore, a 4\% uncertainty on the mass $M_A$, a 6\% uncertainty on
$\alpha$ and a 2\% uncertainty on $a$ combine to give a 8\%
uncertainty on $M_B$. Although in the constrained fit, the value and
uncertainty for $a$ is influenced by the $M_A$ constraint, this
argument for the value and uncertainty of $M_B$ changes little if the values
for $a$ and $\alpha$ from the unconstrained fit are used. 

The apparent photometry corresponding to this preferred fit is given in
Table~\ref{tabPhoto}. The apparent photometry for the system's
combined light comes from Simbad and 2MASS \citep{Cutri03}. The
derived absolute photometry corresponding to individual components of
the system (apportioning equal flux to the two components Aa and Ab)
is also given in this table.
This photometry corresponds to a spectral-type
of $\sim$L5-L7 for GJ~802B \citep{Knapp04}. 

\begin{deluxetable*}{llll}
\tabletypesize{\scriptsize}
\tablewidth{0pt}
\tablecaption{Astrometric Solution for GJ~802AB}
\tablecolumns{2}
\tablehead{\colhead{Parameter} & \colhead{Unconstrained Fit} &
  \colhead{Fixed M$_A$} & \colhead{Constrained M$_A$}}
\startdata
Proper Motion (RA)          &  877.7$\pm$1.0 mas yr$^{-1}$ & 877.7$\pm$1.0 mas yr$^{-1}$&  877.7$\pm$1.0 mas yr$^{-1}$\\
Proper Motion (Dec)         &1722.11$\pm$0.34 mas yr$^{-1}$&1722.2$\pm$1.0 mas yr$^{-1}$& 1722.1$\pm$1.0 mas yr$^{-1}$\\
Parallax                    & 62.2$\pm$1.8 mas             &  63.9$\pm$1.3 mas          &   63.5$\pm$1.3 mas          \\
Orbital semi-major axis ($a$) & 97.5$\pm$5.0 mas             &  92.4$\pm$2.0 mas          &    92.9$\pm$2.1 mas \\
Photocenter semi-major axis  ($\alpha$) & 17.4$\pm$1.2 mas             &   17.2$\pm$0.9 mas         & 17.2$\pm$1.0 mas\\
Epoch (JD)                  & 2453040$\pm$14               &2453028$\pm$15              & 2453031$\pm$13 \\
Period                      &    1105$\pm$ 9 d             &   1105$\pm$ 9 d            &    1104$\pm$ 9 d \\
Eccentricity                &  0.40$\pm$0.08               &0.35$\pm$0.05               &  0.35$\pm$0.05  \\
Argument of periapse        & 90.4$\pm$ 4.3                &  89.0$\pm$ 5.5             &   89.6$\pm$ 4.8 \\
Longitude of ascending node &    21.8$\pm$ 1.4             &  22.7$\pm$ 1.4             &  22.0$\pm$ 1.3  \\
Inclination                 &  83.7$\pm$ 3.0               &  81.1$\pm$ 3.1             &  82.7$\pm$ 3.0 \\
Total Mass                  &  0.426$\pm$0.078M$_\sun$     &0.329$\pm$0.004M$_\sun$     & 0.343$\pm$0.012M$_\sun$\\
Mass of GJ~802Aab            &  0.351$\pm$0.066M$_\sun$     &0.268\tablenotemark{a}      & 0.280\tablenotemark{b}$\pm$0.010M$_\sun$\\
Mass of GJ~802B              &  0.076$\pm$0.013M$_\sun$     &0.061$\pm$0.004M$_\sun$     & 0.063$\pm$0.005M$_\sun$\\
\enddata
\tablenotetext{a}{Fixed from a mass - K magnitude relationship (see text).}
\tablenotetext{b}{Mass with error-bars included as apriori information
  in the Monte-Carlo fits.}
\label{tabOrbit}
\end{deluxetable*}

\begin{deluxetable}{lrrrr}
\tabletypesize{\scriptsize}
\tablewidth{0pt}
\tablecaption{Absolute Photometry for GJ~802}
\tablecolumns{5}
\tablehead{\colhead{Band} & \colhead{m$_A$} & \colhead{m$_B$} & \colhead{M$_{Aa}$(=M$_{Ab}$)} & \colhead{M$_B$}}
\startdata
V\tablenotemark{a} &  14.67 & - & 14.47 &  - \\
J &   9.57$\pm$0.02 &  14.75$\pm$0.27 & 9.34$\pm$0.05 &  13.74$\pm$0.28 \\
H &   9.07$\pm$0.02 &  14.13$\pm$0.09 & 8.83$\pm$0.05 &  13.14$\pm$0.10 \\
K &   8.76$\pm$0.01 &  13.61$\pm$0.08 & 8.53$\pm$0.05 &  12.62$\pm$0.08 \\
\enddata
\tablenotetext{a}{
GJ~802B is assumed to have negligible contribution to V-band.}
\label{tabPhoto}
\end{deluxetable}

\begin{figure}
 \includegraphics{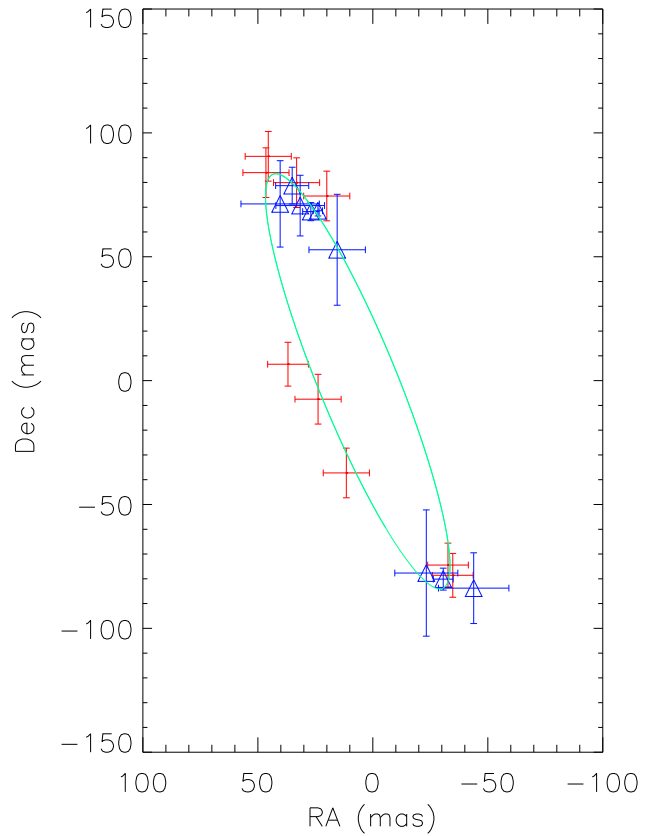}
 \caption{Best fit orbit for GJ~802AB (green), with aperture masking
 points in blue and scaled seeing-limited astrometry in red (see text).}
\label{figOrbit}
\end{figure}

The orbital fit for GJ~802AB is shown in Figure~\ref{figOrbit}, with
the photo-center astrometry measurements scaled by the ratio of total
mass to GJ~802B mass, and with parallax and proper motion removed
so that all points can be plotted at the same scale. This plot also
aids in developing an intuitive feel for our quoted mass error. The
error in the ratio between the mass of GJ~802B and the total mass is
the ratio of the photo-center and orbital semi-major axes (similar to
Equation~\ref{eqnMBIntuitive}): which we claim has a 7\% error.

\section{Comparison with models}
\label{sectModelComp}

As discussed in \citep{Pravdo05}, the activity of GJ~802
(e.g. Figure~\ref{figVisSpectra}), imply that the system is
$\ga$6\,Gyr old.  The kinematics of GJ~802 also imply that the system is old. The
total proper motion of nearly 2 arcseconds per year is 30 times the
parallax, meaning that the tangential velocity of GJ~802 is $\sim$30\,AU
per year, which is $\sim$140\,km\,s$^{-1}$. Based
on the astrometric fit reported in Table~\ref{tabOrbit} and the mean
radial velocity from Tables~\ref{tabInfraredRV} and \ref{tabVisRV2},
we have calculated the space motion of GJ~802 using the {\tt gal\_uvw}
routine from the {\tt IDL} Astronomy User's 
Library\footnote{http://idlastro.gsfc.nasa.gov/}. This routine is in
turn based on the mathematics in \citet{Johnson87}. The $(U,V,W)$
 space motion calculated to be $(134,-51, 21)$\,km\,s$^{-1}$, which is 
$(134,-39,28)$\,km\,s$^{-1}$ with respect to the local standard of
rest. The lag in $V$ and very high $U$ space velocity is quite
inconsistent with the thin-disk population. For example, the {\em
  HIPPARCOS} sample of nearby early M stars has a $U$ dispersion of
32\,km/s, and the sample earlier than spectral-type F5, representative
of a $\sim$1\,GYr old population, has a $U$ dispersion of only
22\,km\,s$^{-1}$ \citep{Mignard00}.  The 39\,km\,s$^{-1}$ $V$ lag of GJ~802 
is unusually small for a Population~II halo star, but is consistent
with the thick disk \citep{Casertano90}. GJ~802 is very unlikely to be
a runaway star that is confused with the thick disk population,
because any dynamical interaction capable of giving it a $\sim
100$km\,s$^{-1}$ peculiar velocity would also break the wide binary
GJ~802AB apart. Therefore, these kinematic properties place GJ~802 at
an age of $\sim$10\,GYr, and almost certainly older than 3\,GYr
\citep{Bensby03}. 

\begin{figure*}
\epsscale{1.00}
\plotone{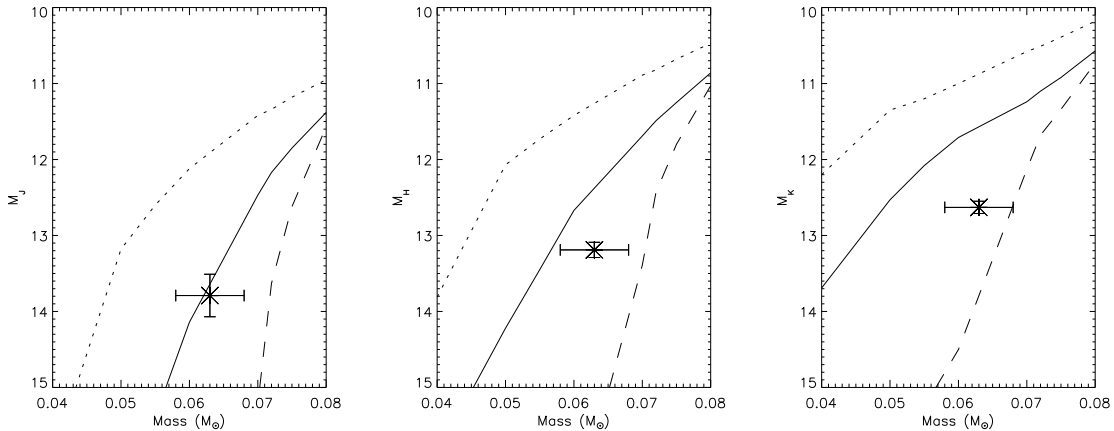}
\vspace{1cm}
\caption{Absolute magnitudes of GJ~802B compared to the DUSTY models
  at 0.5, 1 and 5 GYr. The comparison is for J-band (left panel),
  H-band (center panel) and K-band (right panel). Model ages are
  0.5\,GYr (dotted line), 1\,GYr (solid line) and 5\,GYr (dashed line).} 
\label{figModelComp}
\end{figure*}

We can compare these ages to the ages derived from the modelling of
GJ~802B.
Figure~\ref{figModelComp} shows the absolute magnitudes of GJ~802B
compared to the DUSTY models of \citet{Baraffe02}. The colors of these
models are clearly too red, a well known property of these models
for objects that are of mid-L or later spectral type
\citep{Baraffe03}. The predicted age is between 1 and 5\,GYr in
age in all bands. We chose not to 
plot the results from the COND models from the same group
\citep{Baraffe03} because these models are much too blue and really
only applicable to T dwarfs.


Although these models clearly do not represent GJ~802B, mostly due to
difficulties in opacity treatment and dust in brown dwarf atmospheres,
it is expected that luminosity and effective temperature are a much
more certain prediction of brown dwarf evolutionary models than fluxes
in specific bands \citep{Burrows93,Chabrier00}. 
A better comparison for the GJ~802B photometry is therefore possible 
with the models of \citet{Burrows06} that correctly model the basic
colors of L dwarfs and the L to T transition.

%
At log($g$)=5.2 (appropriate for a $\sim$0.06\,$M_\sun$ brown dwarf),
the best fitting model from \citet{Burrows06} has absolute magnitudes
$M_J =14.05$, $M_H =13.10$ and $M_K = 12.78$. This model has 
$T_{\rm eff} = 1400$\,K with a total luminosity of 
$2.9 \times 10^{-5}$\,L$_\sun$.
This model 
Luminosity and temperature can now be compared with evolutionary
models, placing a 0.063\,$M_\sun$ GJ~802B at 1.9\,GYr according to the
\citet{Burrows97} track. GJ~802B would only
be consistent with an old (i.e. $>$5\,GYr) object if its mass were
0.07\,$M_\sun$, which is inconsistent with our
dynamical mass at $1.5\sigma$.

\section{Conclusions and Discussion}
\label{sectConclusion}

GJ~802 is a triple system with component masses of $\sim$0.14, 0.14
and 0.063\,$M_\sun$. The inner pair of equal-mass stars, GJ~802Aab,
has an orbital period of 0.795 days and an 
inclination between 74 and 83 degrees. The outer component,
GJ~802B, has a $\sim$3 year orbit and a mass of $0.063 \pm
0.005$\,M$_\sun$. The inclination of the GJ~802B orbit, $\sim$83
degrees, is consistent with co-planarity.
The most promising way to decrease the mass
uncertainty would be to obtain an accurate radial velocity curve of
the 3 year orbit, with an expected amplitude of $\sim$3\,km/s.

The brown dwarf ``desert'' is generally defined as a lack of low-mass
companions around solar-type stars, but a similar lack of very unequal
mass-ratio companions has been found around very low-mass stars
\citep{Close03}. However, as a triple system, GJ~802 no longer fits in this
category. In fact, there are a large number of brown dwarfs known in
triple or higher order multiple systems. This entirely consistent
with dynamical star-formation simulations \citep{Delgado04}, but may
also relate to the details of the fragmentation process as a cloud
with relatively high angular momentum collapses. Measuring the
mutual inclination (i.e. degree of co-planarity) of this and other
low-mass triples with long-baseline interferometry could help to
determine the mechanism responsible for producing triples like GJ~802.

The kinematics of GJ~802B place it in the thick disk, at an age of
$\sim$10\,GYr, while the model-derived age for GJ~802B is
$\sim$2\,GYr. Although this discrepancy is only significant at the
93\% confidence level so far (1.5$\sigma$), we will list several
possibilities for reconciling the discrepancy: 1) GJ~802Aab could actually
have a total mass of $>$0.30\,$M_\sun$, consistent with the astrometry, but
be underluminous by $\sim30$\% in K-band compared to field
stars. GJ~802B would then be placed right on the substellar boundary
at $\sim$0.07\,$M_\sun$, a naively unlikely position given the very
small sample of astrometric STEPS binaries from which GJ~802 was
taken \citep{Pravdo05}.
2) GJ~802Aab could have an apparent total mass of $>$0.30\,$M_\sun$ because
of an additional low-mass component in a $\sim30$\,day, just stable
orbit. This could also increase our dynamical mass of GJ~802B to
$\sim$0.07\,$M_\sun$. 
3) GJ~802 could actually be $<$2\,GYr old but have experienced a
unique dynamical past that 
gave it a high space velocity without tearing the wide binary
apart. Or, 4) models for old brown dwarfs are systematically under-predicting
luminosities. These model errors could relate to, e.g. the effects of
magnetic fields that can hinder heat flow in brown dwarfs \citep{Chabrier07}.
Given that GJ~802B is currently the only $\sim$Gyr or
older brown dwarf with an accurate dynamical mass, we find this fourth
possibility most likely: a hypothesis that will be testable within the
next few years as more field brown dwarfs have accurate mass
determinations. 

\acknowledgements

M.I. would like to acknowledge
Michelson Fellowship support from the Michelson Science Center and the
NASA Navigator Program. A.L.K. is supported by a NASA/Origins grant to
L. Hillenbrand. This work is partially supported by the National
Science Foundation under Grant Numbers 0506588 and 0705085. 
Some of the data presented herein were obtained at the
W.M. Keck Observatory, which is operated as a scientific partnership
among the California Institute of Technology, the University of
California and the National Aeronautics and Space Administration. The
Observatory was made possible by the generous financial support of the
W.M. Keck Foundation. The authors wish to recognize and acknowledge
the very significant cultural role and reverence that the summit of
Mauna Kea has always had within the indigenous Hawaiian community.  We
are most fortunate to have the opportunity to conduct observations
from this mountain. Based partly on observations obtained at the Hale
Telescope, Palomar Observatory, as part of a collaborative agreement
between the California Institute of Technology, its divisions Caltech
Optical Observatories and the Jet Propulsion Laboratory (operated for
NASA), and Cornell University. Based partly on observations obtained 
at the Palomar 60-inch robotic telescope.

Facilities: \facility{Keck:II (NIRC2)}, \facility{Hale (PHARO,EAE)}, \facility{PO:1.5m}

\end{document}